\documentclass[referee]{aa} 
\usepackage{graphicx} 
\begin{document} 
   \title{The LEDA galaxy distribution : I. Maps of the Local Universe} 


   \author{H. Courtois\inst{1},
 G. Paturel\inst{1},
           T. Sousbie\inst{1}, 	    
	F. Sylos Labini\inst{2}
       }

   \institute{Centre de Recherche Astronomique de Lyon (CRAL), 
              9 avenue Charles Andr\'e, 69561 Saint Genis Laval Cedex, France 
              \email{courtois@obs.univ-lyon1.fr}
   \and{Laboratoire de Physique Th\'eorique, Universit\'e Paris XI, Batiment 211,
91405, France } 
             } 
  \offprints{H. Courtois}

   \date{Received November 30, 2003; accepted february 2004, in press}

   \abstract{ 
In order to investigate the properties of large-scale structures of galaxies in the 
universe, we  present an analysis of their spatial distribution at 
z$<$0.033. We used the LEDA extragalactic database containing over 1 million of 
galaxies covering the all-sky and the SDSS data included in the public release DR1, 
yielding to a sample of around 134,000 galaxies having a measured 
redshift in two survey areas representing 690 sq. degrees. The results of the study 
are 2D, 3D maps
 and magnitude number counts of galaxies, drawn
from B-band samples.    
   } 
\authorrunning{H. Courtois et al. }
\titlerunning{Maps of the Local Universe }
   \maketitle 
%
%
%
%
%




\section{Introduction} 

Angular and radial maps of the Local Universe
are the first step for astronomers to investigate
the nature of large scale structures observed
in the galaxy distribution.
An early discovery of a large scale structure
was the "great wall" of galaxies, in the CfA1 
redshift survey  \cite{CFA} and \cite{CFA1}.  Building on this promising
technique for tracing 3-D structures,
deeper redshift surveys including: CfA2 \cite{CFA2}, IRAS,
LCRS, 2dF \cite{2DF}, and SDSS \cite{DR1}, all show new more extensive
and complex structures such as walls, chains,
voids, and superclusters.  A related approach
to quantify the complexity of the galaxy distribution
is to construct the galaxy number counts
in mag, N(m), and in distance, N(r), using
statistical indicators: spatial and angular
two-point correlation functions.  Combining approaches,
physical interpretation of these counts leads to the 
understanding of the nature of large scale structure,
and to its cosmological implications for
the models of formation of galaxies.  Much attention
is focused on the determination of the scale of
homogeneity in the universe.  It is not a trivial
concern, some statistical indicators are interpreted
as showing the scale of homogeneity to be as small
as 15 Mpc; while the new redshift surveys show extended 
structures larger than 100 Mpc, dwarfing even cluster sized 
stucture.
In this present paper, we will not directly address this issue,
but instead focus on the nearby Universe, creating
a global picture for the redshift and number count
distribution of galaxies.

The paper is organized as follows : section 2 presents the data, the completness studies,
the number counts and
the construction of volume limited subsamples; section 3 shows the 2D and 3D maps of the Local Universe
in various coordinates systems and preliminary comparison with SDSS DR1 
and section 4 gives our conclusions.


\section{Building the  samples} 

\subsection{The LEDA database}
   \begin{figure} 
   \centering 
   \includegraphics[width=8cm]{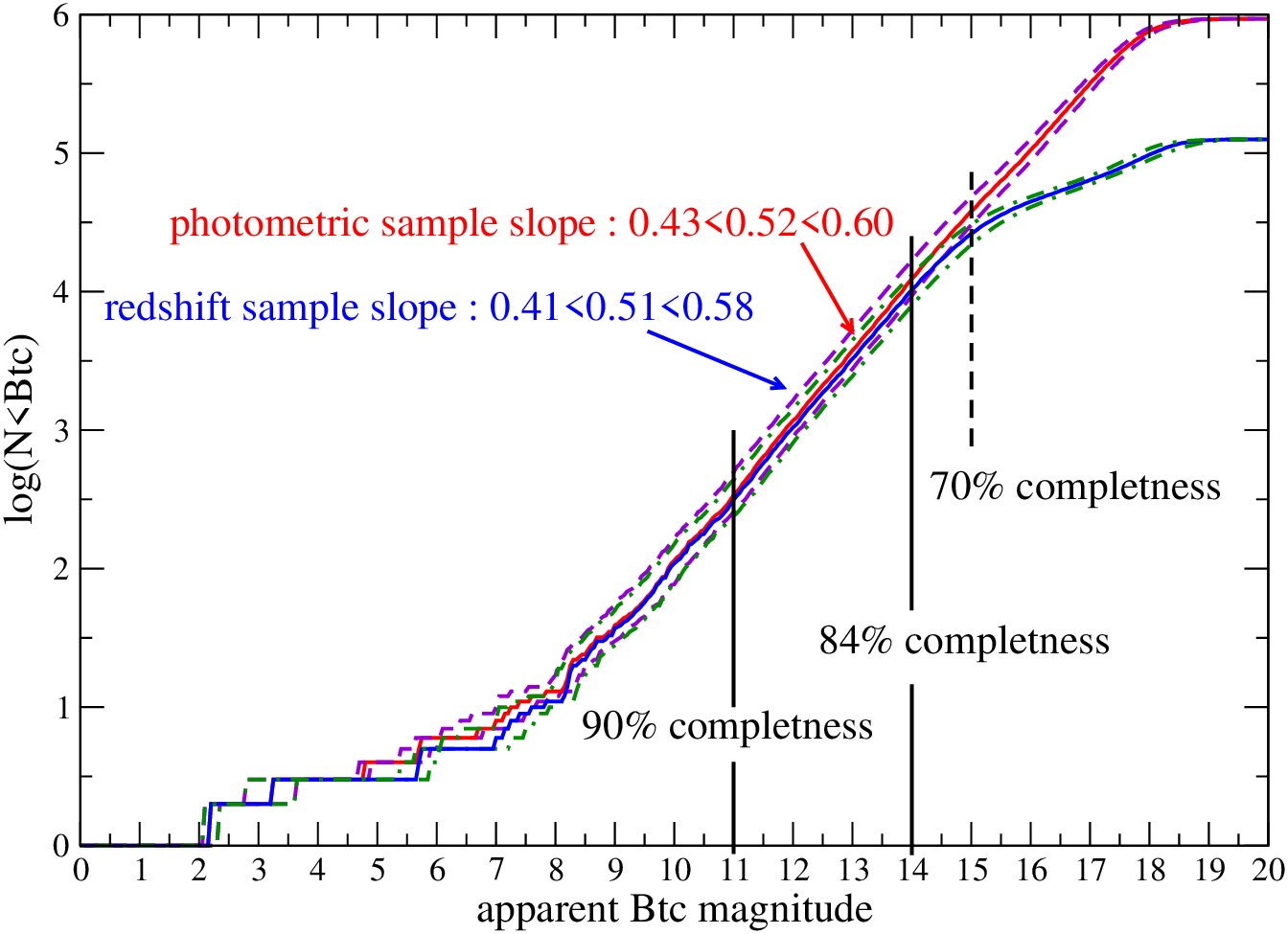} 
      \caption{
Magnitude number counts for LEDA : decimal logarithm of the cumulative
number of observed galaxies versus the extinction corrected B total magnitude (Btc). The zone of avoidance (-15deg$<$galactic latitude$<$15deg) has been cut
out of the catalog, and its completness is not tested here.
There is a clear change of slope at Btc=14.5 in the redshift catalog and at
Btc=17 in the photometric one. These indicate the limit in magnitude where the
catalogs hit incompletness.The red and green dashed lines show the smallest and largest
error bars on the slope due to the error measurements on Btc.Their slopes corresponds to
the largest and smallest values given for the mean slope. } 
         \label{figNmLEDA} 
   \end{figure} 
%
The Lyon-Meudon Extragalactic Database : LEDA, was the first database created
in order to collect all published measurements on galaxies related to studies
of the structure and kinematics of the Local Universe. On this purpose, the database
is limited to galaxies closer than 60,000 $km.s^{-1}$. It was created by Georges Paturel
in 1983 (see \cite{Patu97}) relaying the work of G\'erard De Vaucouleurs and his famous
RC1 and RC2 catalogs (\cite{RC12}). The RC3 catalog (\cite{RC3})
was made using the LEDA database. 
This archive has stimulated a large and varied number of studies, 
over 100 refereed papers, that address large scale structure 
and peculiar velocity flows.
 In the year 2000, the
database galaxy number was increased to over 1 Million. And for the 20th anniversary
of LEDA in 2003, completely new studies of the all-sky Local Universe are possible.
One should insist on the fact that there is no other possibility to work with
all-sky homogeneized magnitudes in a volume of $3.10^{6} Mpc^{3}$.
 This is true for B selected surveys, however the 2MASS
survey, was designed precisely to be
a homogenized , allsky, uniform survey of galaxies
(as well as the Milky Way).  In the NIR, 2MASS
is more complete than optical survey since it
penetrates the Zone of Avoidance.  One can see
the spectacular results in the allsky data release
(XSC: extended source catalog, see \cite{jarrett1} and \cite{jarrett2}).

 In all the paper
we use H$_{0}$=100 Mpc/km/s. The non-Beta-version 
of the DR1 of SDSS has been released in June 2003,
and is simultaneously employed to discern the large scale structures in a slice 
of 600Mpc depth. The extinction correction on LEDA B magnitudes is now made
according to \cite{Schlegel} since 2002 in LEDA. Hereafter we refer
to these B total magnitudes corrected for galactic extinction as Btc magnitudes.

   \begin{table} 
      \caption[]{Data in LEDA in 2003} 
         \label{LEDA} 
         \begin{tabular}{ll} 
            \hline 
            \noalign{\smallskip} 
            data     &  number \\ 
            \noalign{\smallskip} 
            \hline 
            \noalign{\smallskip} 
            galaxies & 1,316,143     \\ 
          B magnitudes  & 929,815            \\ 
          radial velocities   & 213,446 \\ 
          HI width measurements    & 51,621            \\ 
            \noalign{\smallskip} 
            \hline 
         \end{tabular} 
   \end{table} 

\subsection{Completness studies}
In order to build representative maps of the Local Universe, 
we should first test the completness of the samples. In the case of the LEDA 
database, we compare the number 
counts in apparent magnitude in the B band. 
The magnitudes are corrected from galactic extinction, as notified before, and thus noted hereafter 
Btc. In the database there are about 930,000 galaxies with a measured B magnitude (hereafter photometric
catalog), while 
125,685 of them have also a measured redshift (hereafter redshift catalog).

Before testing the completness
of the catalogs available in LEDA, we cut out the Zone of Avoidance
(-15deg $<$ galactic latitude $<$ 15 deg).  The Zone is woefully
incomplete, so one would expect many galaxies
to be missing (as well as galaxy clusters and
superclusters, including the Great Attractor).
Galactic extinction is accounted for in LEDA Btc magnitudes, but that
really only works for the low extinction cases
(Av $<$ 1). In the Galactic Plane the extinction
is very patchy and is not always well traced
with the Schlegel maps. So it is better for our studies of completness
to cut this zone out.

On Figure \ref{figNmLEDA} we plot the magnitude number 
counts of the two subsamples : the "photometric" one and the "redshift" one. One 
can see that there is a linear growth of the number of galaxies with the tested 
volume. However the slopes are not following the 0.6m expected if the distribution 
of galaxies was homogeneous in the Local universe. For the photometric sample, 
the slope is 0.52 and for the redshift one it is 0.51. These slopes are measured 
only in the part where the incompleteness can be neglected.
If we suppose that the photometric catalog is complete, then the redshift one
contains  84\% of the galaxies at Btc=14.5 and 90\% at Btc=11.

In the literature there 
are various interpretations for the values of these slopes : local underdensity, 
fractal behaviour below the scale of homogeneity, incompletness of the low surface brightness 
galaxies. The studies of the SDSS data which are much deeper, showing a 
slope of 0.45 as we will see in a following paper, tends to cancel out the local 
underdensity explanation, or at least extend this local underdensity to scales of 
600 Mpc depth. Recent surveys dedicated to low surface brightness galaxies tend to 
show that those galaxies are not present in large-scale voids nor in our vicinity. 
We can then neglect their effect on the magnitude number counts slopes.  
The departure from linearity of these slopes gives us the limit of completness of 
the samples. It is Btc=14.5 for the redshift catalog and Btc=17 for the photometric 
one. Knowing those limits we will be able to cut volume limited samples.

We did various tests to quantify the error on the magnitude number count slope due to the errors
on the magnitudes. One can find previous studies of LEDA galaxy number counts in 
\cite{DiNella96} and \cite{Patu94}. It is not the purpose of this paper to discuss deeply
those counts, here we use them only in order to build representative samples of galaxies
to allow a construction of a set of maps. 
We  measured the slope in 3 cases taking into account an error on the magnitudes 
 which is respectively zero, equal to the real error and the double of the real error.
In this manner we check that the error has low influence on the slope of the magnitude number
counts.On Figure \ref{figNmLEDA} we wrote the smallest and the largest values of the slopes,
we could obtain when accounting for the maximum of the error bars on each value of
the magnitude.The errors on the magnitudes in LEDA are so large, that the error on the slope
is about 0.09, so we can not totally exclude a slope of 0.6, in particular for the photometric catalog.

\subsection{Volume limited subsamples}

   \begin{figure} 
   \centering 
   \includegraphics[width=8cm,angle=90]{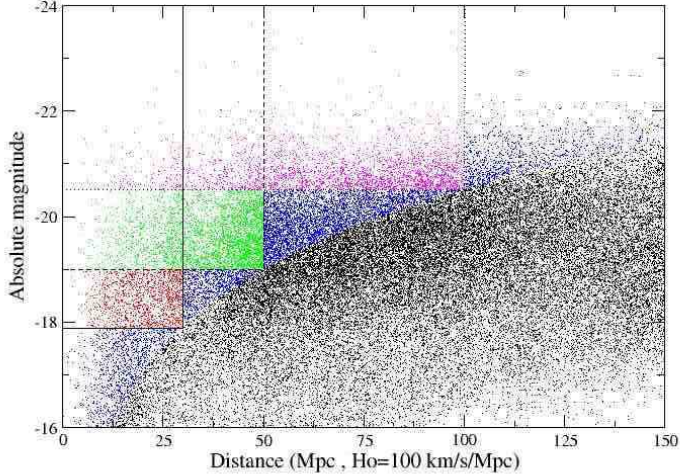} 
      \caption{
Absolute magnitude limits for the volume limited subsamples.The redshift
catalog shown in black contains 125,685 galaxies.
The blue points, 16,906 sources, denote a 
Btc less than 14.5 selection. The red points, 2504 sources, denote a Volume 
Limited (VL) cut at 30 Moc; the green points, 4434 sources, denote 
a VL cut at 50 Mpc; purple points, 2814 sources, denote a VL cut at 
100 Mpc. } 
         \label{Figcomp1} 
   \end{figure} 
%

\begin{figure} 
   \centering 
   \includegraphics[width=8cm,angle=90]{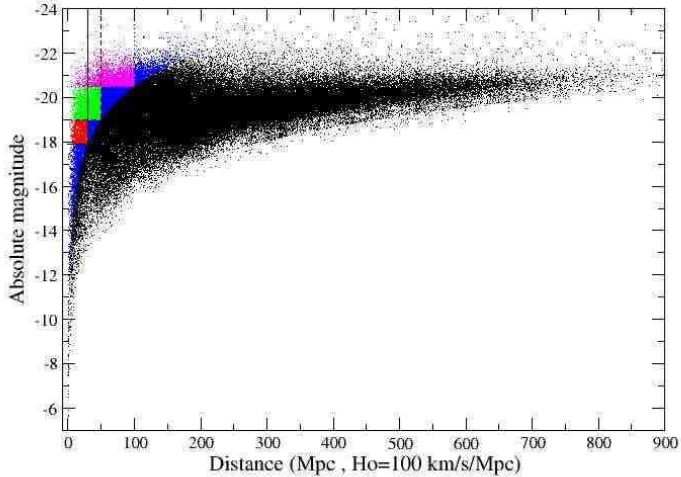} 
      \caption{Completness limits for the volume limited subsamples. The colours
are the same as in figure\ref{Figcomp1}. } 
         \label{Figcomp2} 
   \end{figure}

On Figures \ref{Figcomp1} and \ref{Figcomp2} we can see how the cut in apparent magnitude 
Btc $<$ 14.5 affects the redshift samples. From 125,685 galaxies (in black) we have now (in blue) 
only 16,906 galaxies left. Those samples have been observed with selections in apparent magnitude,
 thus they contain a wide variety of galaxies. The distribution in absolute magnitude within these 
samples is incomplete. They contain dwarf galaxies, normal galaxies and giant galaxies, but each
 of these classes is not complete. The apparent magnitude surveys do not contain all dwarf, all 
normal and all giant galaxies. In order to select only galaxies statistically representative, 
ie,  having  the same absolute magnitude, we apply a cut in distance and in absolute magnitude. 
We build three volume limited subsamples at 30 (red), 50 (green) and 100 (purple) Mpc.
 The number of galaxies left in those samples can be read from the table \ref{tableVL}. As we go deeper 
 with the volume limited samples, we select only luminous galaxies.
However, one can see on figure
\ref{Figcomp2} that the resulting number of galaxies in the deepest
volume limited subsample at 100 Mpc contains only about 2\% of the
initial redshift catalog of 125,685 galaxies.

%

   \begin{table} 
      \caption[]{Volume limited subsamples} 
         \label{tableVL} 
         \begin{tabular}{ccc} 
            \hline 
            \noalign{\smallskip} 
            lim. distance     &  lim. abs. magnitude & number of gal. \\ 
            \noalign{\smallskip} 
            \hline 
            \noalign{\smallskip} 
          30 Mpc & -17.9  & 2,504 \\ 
          50 Mpc   & -19  &  4,434 \\ 
	100 Mpc   & -20.5 & 2,814 \\
            \noalign{\smallskip} 
            \hline 
         \end{tabular} 
   \end{table} 

\section{Maps of the Local Universe}
\subsection{2D maps}

\begin{figure*} 
   \centering 
   \includegraphics[width=14cm,angle=90]{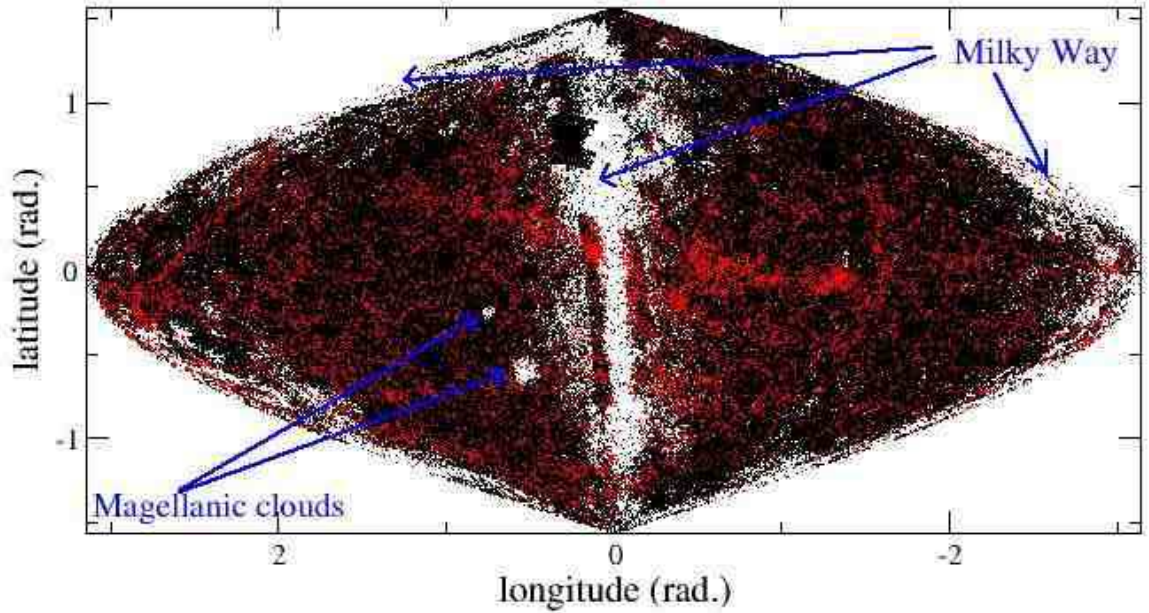}
   \caption{2D map of the Local Universe : in black : all galaxies (983,300) with a measured
B magnitude in LEDA
     and in red : the magnitude limited sample: Btc $\le$ 14.5 (21,859 galaxies).
  Flamsteed equal area projection in supergalactic coordinates.}
              \label{Flamall}%
    \end{figure*}

\begin{figure*} 
   \centering 
   \includegraphics[width=14cm,angle=180]{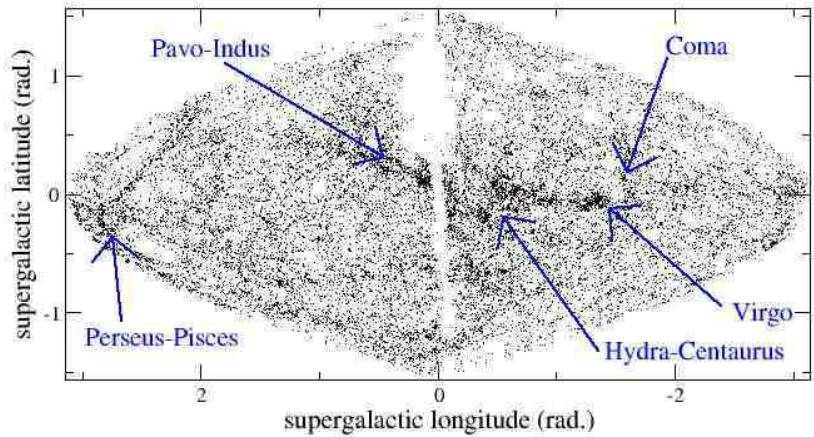} 
   \caption{2D map of the Local Universe : All galaxies (21,859) where 
  Btc $<$ 14.5 in LEDA are the black dots. Major superclusters in our neighborhoods
are shown. Flamsteed equal area projection in supergalactic coordinates} 
              \label{FlamBtc}
    \end{figure*}

\begin{figure} 
   \centering 
   \includegraphics[width=10cm]{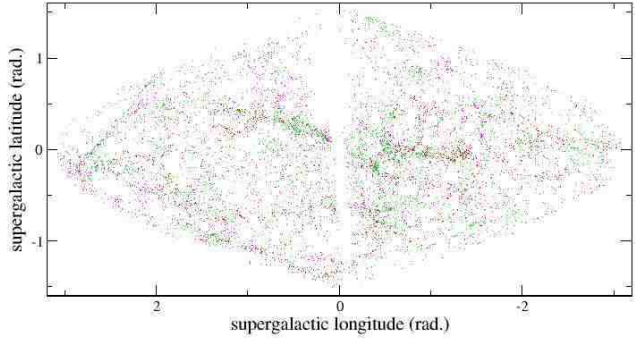} 
   \caption{2D map of the Local Universe : Volume limted subsamples.
Flamsteed equal area projection in supergalactic coordinates.One can see that the
different VL subsamples don't trace the same structures.  Colour code is : red VL30 Mpc,
green VL50 Mpc, purple: VL100Mpc.} 
              \label{FlamVL}
    \end{figure}

On Figures \ref{Flamall}, \ref{FlamBtc} and \ref{FlamVL}  we show angular maps of the Local Universe. 
They are Flamsteed projections in supergalactic coordinates. 
This method of representing an angular catalog 
has the advantage to be an "equal area projection", thus conserving the projected areas. 
But of course, as any projection, there are deformations effects near the poles 
(up and bottom of the figures) and all along the sides. 

On Figure \ref{Flamall}, the catalog 
of all galaxies having a measured B magnitude is shown in black. The galaxies
having a magnitude brighter than 14.5 are in red. We can see that only this first 
selection step allows to distinguish between real structures and incompletness effects. 
For example near the zone of avoidance, in the center of the figure, one can see 
two vertical bands corresponding to the Las Campanas Redshift Survey (LCRS). These 
bands do disappear after the cut in apparent magnitude, ie this survey has a 
completness limit much deeper  than the completness limit of the  composite all-sky 
collection of redshift surveys in LEDA.

One structure that can be seen very clearly 
on Figure \ref{FlamBtc} is a kind of wave in the structures appearing in the upper left part of 
the figure and in the lower right part. Such a deformation in a 2D map can be due 
to a large scale plane of galaxies in the 3D distribution. As a matter of fact,  
this wave is passing through  Perseus-Pisces, Pavo-Indus, Hydra-Centaurus, Virgo 
and Coma superclusters.
It is known and described as
the hypergalactic plane in \cite{DiNella94}.
The hypergalactic plane is not equivalent to the supergalactic plane.
The supergalactic plane describes the local supercluster (LSC) plane at a shorter distance from the observer. 

In fact the hypergalactic
plane contains the local supercluster as Virgo cluster (the center of the LSC)
is lying on this plane extending to larger scale.

\subsection{3D maps}

\begin{figure} 
   \centering 
   \includegraphics[width=8cm,angle=90]{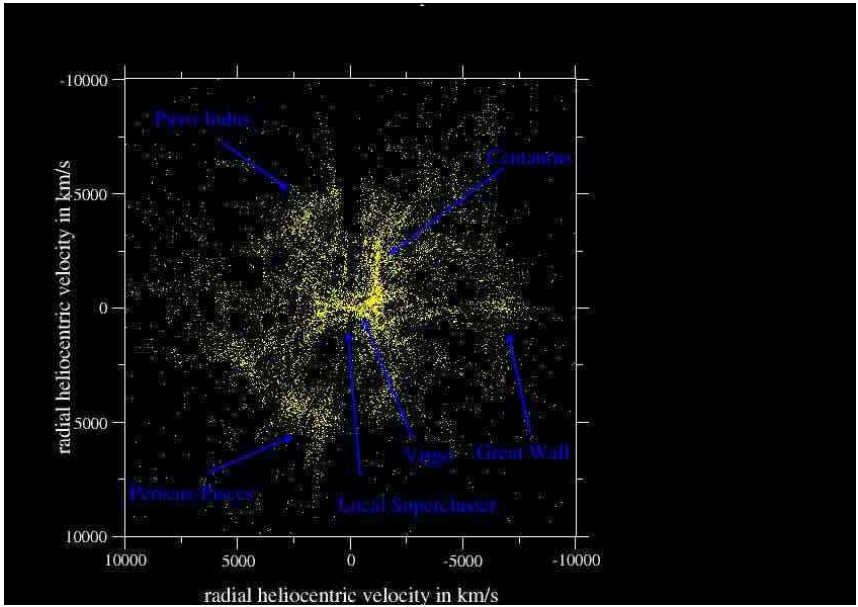} 
   \caption{View face-on of the XY plane in supergalactic coordinates, slice of 50 Mpc along the Z axis.} 
              \label{FigXYSGVL}%
    \end{figure}

\begin{figure} 
   \centering 
   \includegraphics[width=8cmi,angle=90]{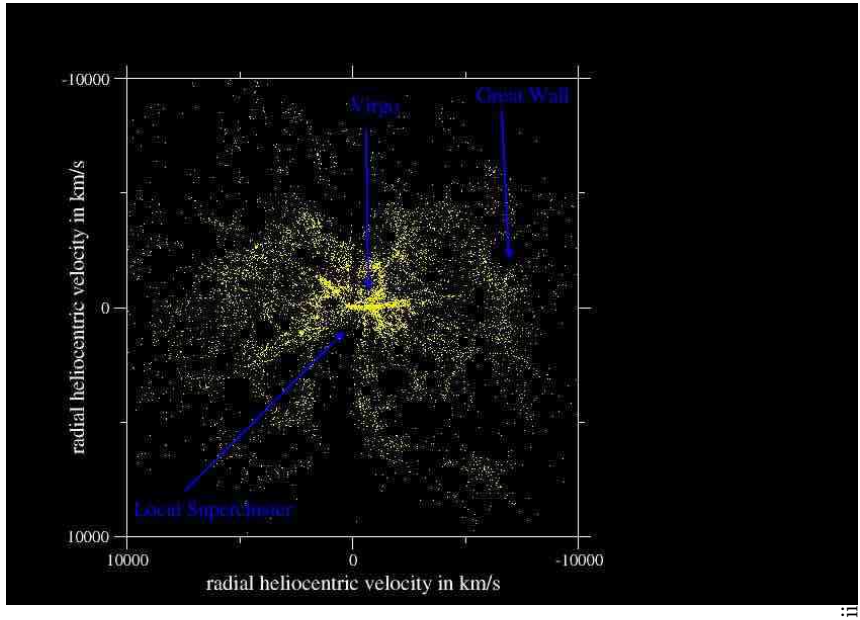} 
   \caption{View face-on of the XZ plane in supergalactic coordinates, slice of 50 Mpc along the Y axis.} 
2A
              \label{FigXZSGVL} 
    \end{figure}

After inspection of the galaxy distribution using 2D maps, 
it is constructive to view the allsky 
maps with a 3-D projection using the VL redshift sample. Here we project the 
supergalactic coordinates and redshift into an XYZ volume;
 figures \ref{FigXYSGVL} and \ref{FigXZSGVL}.
The latest largest all-sky 3D maps where published almost ten years ago (\cite{DiNella94}, 
\cite{DiNella95}, \cite{DiNella97}, \cite{Amendola97}),
this paper improves the data number and more importantly the quality of
the magnitudes used.   


\subsection{Preliminary comparison with SDSS DR1}

\begin{figure} 
   \centering
   \includegraphics[width=10cm]{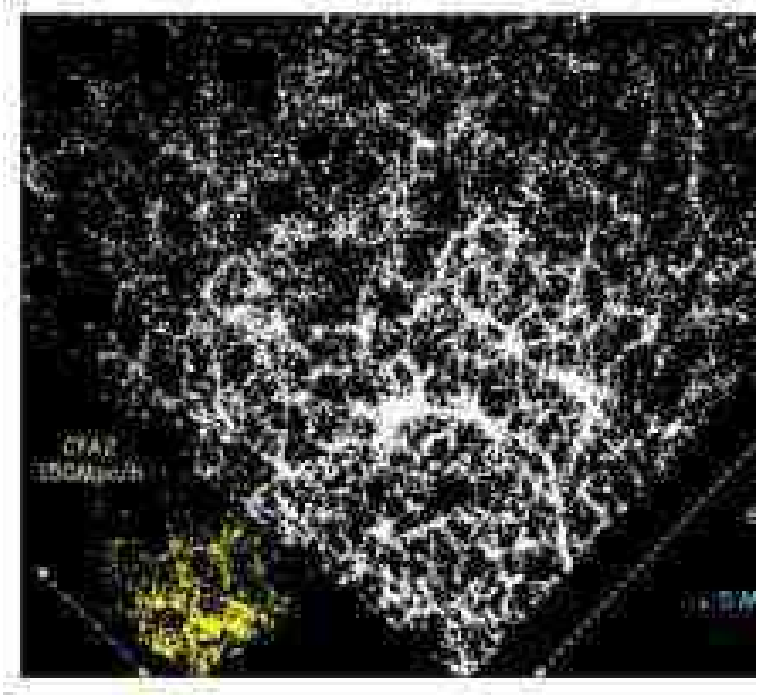} 
   \caption{
   Comparison between the first public data release of SDSS and CFA2  } 
              \label{FigSDSS} 
    \end{figure}

To go further in the mapping of the Local Universe, every astronomer working on 
large-scale structures has been waiting for Sloan Digitized Sky Survey (SDSS). 
These data are of first quality order because they provide the magnitudes in 5 bands, 
which will be very helpfull to quantify the evolution effects on the number counts.
On Figure \ref{FigSDSS}, one can see the first public datarelease (DR1) of the Sloan Digitized 
Sky Survey (SDSS) \cite{DR1}. The first amazing fact to note is the continuity of  bubbles, 
walls, and voids sequence up to a scale 6 times larger than the one we just 
discussed : up to z =0.2. The second point is the structures {\it  size} : we can see 
a chain (or a wall ??) of galaxies extending on about 300 Mpc. As a comparison, 
we ploted the CfA2 (\cite{CFA1} and \cite{CFA2})redshift survey containing 
the Great Wall of galaxies, which 
is about 100 Mpc long. We are now seeing a structure 10 times larger than the Great 
Wall..
The 2dF redshift survey (\cite{2DF}) probed a region with comparable 
depth to SDSS, revealing large scale structure with equally impressive detail 
and complexity. Combined, these redshift surveys suggest structures as large 
as 300 Mpc in extent, probably associated with "walls" of galaxy clusters.

To help visualize structure in the Local Universe, we have created 3 animated 
renderings of the LEDA and SDSS redshift samples in XYZ space. The first, LEDA.avi, 
shows all LEDA galaxies with redshifts measurements; the second, LEDA145.avi, shows the 
LEDA galaxies brighter than 14.5 mag in Btc; and the third, LEDASDSSDR1.avi, 
combines the LEDA redshifts with those of SDSS (DR1). The movies show the galaxy 
distribution in 3-D space, rotated about each axis separately. For the LEDA/SDSS 
movie, the LEDA points are colored blue, while the SDSS points are colored yellow. 
Note the much greater depth, but more patchy coverage, of the SDSS sample 
compared to the LEDA redshift sample.

It is striking to see 
that the very local view of the local universe with bubbles, chains, walls 
can still be seen at a scale 10 times larger. One cannot guess the intrinsic 
scale of  the structures by looking at these maps. We can really begin to speak 
about hierarchy in the large scale structures, another word for scale invariance. 
Studies of the statistical properties of such distributions are ongoing and we will 
present them very soon. For the moment, it is for instance clear  that a galaxy 
correlation length of 15 Mpc, could hardly be a  meaningfull statistical indicator 
tracing the galaxy clustering properties  in the Local Universe.

\section{Conclusions} 
With the astonishing large scale structure observed by 2MASS, SDSS, 2dF 
and through the LEDA archive, it is increasingly clear that the cosmological 
models must account for structures ten times larger than previous convention, 
upwards of 300 Mpc.
The second idea is that we cannot 
simply estimate the scale at which one is looking at the structures, just by looking
at the maps : we are 
seeing similar structures at different scales. This could be another argument 
in favor of using general tools that can describe both homogeneous and scale 
invariant distributions, to analyse the galaxy point distribution.

\begin{acknowledgements} 
We thanks very much the referee
Thomas Jarrett for his help in improving this paper.
\end{acknowledgements}

\end{document}